\documentclass[pdflatex,sn-mathphys-num]{sn-jnl}
\usepackage{setspace}%
\usepackage{graphicx}%
\usepackage{multirow}%
\usepackage{amsmath,amssymb,amsfonts}%
\usepackage{amsthm}%
\usepackage{mathrsfs}%
\usepackage[title]{appendix}%
\usepackage{xcolor}%
\usepackage{textcomp}%
\usepackage{manyfoot}%
\usepackage{booktabs}%
\usepackage{listings}%
\usepackage[ruled]{algorithm2e}
\usepackage{arydshln}
\usepackage{hyperref}
\usepackage[mathscr]{euscript}
\usepackage{subcaption}
\theoremstyle{thmstyleone}%
\newtheorem{theorem}{Theorem}
\newtheorem{definition}[theorem]{Definition}% 
\theoremstyle{thmstyletwo}%
\newtheorem{example}{Example}%
\newtheorem{remark}{Remark}%

\theoremstyle{thmstylethree}%

\begin{document}

\title[Article Title]{Probabilistic State Estimation of Timed Probabilistic Discrete Event Systems via Artificial Neural Networks [Draft Version].}

\author*[1]{\fnm{Omar Amri}}\email{omar.amri@univ-lehavre.fr}

\author[2]{\fnm{Carla Seatzu}}\email{carla.seatzu@unica.it}

\author[2]{\fnm{Alessandro Giua}}\email{giua@unica.it}

\author[1]{\fnm{Dimitri Lefebvre}}\email{dimitri.lefebvre@univ-lehavre.fr}

\affil*[1]{\orgdiv{GREAH}, \orgname{Université Le Havre Normandie}, \orgaddress{\street{75 Rue Bellot}, \city{Le Havre}, \postcode{76600},\country{France}}}

\affil[2]{\orgdiv{DIEE}, \orgname{University of Cagliari}, \orgaddress{\city{Cagliari}, \postcode{09124}, \country{Italy}}}


\abstract{This paper is about the state estimation of timed probabilistic discrete event systems. The main contribution is to propose general procedures for developing state estimation approaches based on artificial neural networks. It is assumed that no formal model of the system exists but a data set is available, which contains the history of the timed behaviour of the systems. This dataset will be exploited to develop a neural network model that uses both logical and temporal information gathered during the functioning of the system as inputs and provides the state probability vector as output. Two main approaches are successively proposed $(\textit{\textbf{i}})$ state estimation of timed probabilistic discrete event systems over observations: in this case the state estimate is reconstructed at the occurrence of each new observation; $(\textit{\textbf{ii}})$ state estimation of timed probabilistic discrete event systems over time: in this case the state estimate is reconstructed at each clock time increment. For each approach, the paper outlines the process of data preprocessing, model building and implementation. This paper not only proposes groundbreaking approaches but also opens the door to further exploitation of artificial neural networks for the benefit of discrete event systems.}


\maketitle
\doublespacing
\section{Introduction}
State estimation is a fundamental question in systems and control theory. It has an important role in the comprehension, analysis, and control of dynamic systems. It becomes a challenging task when dealing with a system, whose initial state may not be exactly known, or (and) whose behaviors are non-deterministic \cite{SHU20083054}. As far as Discrete Event Systems (DES) are concerned, Petri nets, finite automata and their extensions have been generally exploited to deal with the problem of DES state estimation.

For systems represented as finite automata, the problem of state estimation is initiated by Ramadge et al. \cite{Ram1} and Caines et al. \cite{Caines}, where the concept of observability is introduced and the observer structure is developed. Ozveren et al. \cite{Ozveren} proposed an approach for constructing an observer that reconstructs the state of finite automata after observing a word of bounded length. In a timed setting, Gao et al. \cite{Gao1} propose a region automaton and a $\lambda-$observer for current state estimation over time under no event observation, where the state estimate is reconstructed as time passes without any observation being recorded. Li et al. \cite{Li} propose a timed observer that uses the time stamps of observations to refine the state estimation. Lai et al. \cite{Lai} convert time to weight to construct the observer. Shu et al. \cite{Shu1} have investigated the problem of state estimation of probabilistic discrete event systems. An attempt to address state estimation within a context that incorporates both timed and probabilistic aspects was done by Lefebvre et al. \cite{jdeds1}, \cite{jdeds2}. The system is modeled using labeled timed probabilistic automata, defined as a special type of continuous time Markov models. In these works, observations and their time occurrence are used to refine the state estimation. These results are extended in \cite{dim3} to characterize two main cases,  the case where the silent closure\footnote{A silent closure is the period of time where the system remains silent i.e., no observation is produced.} is finite and the case where the silent closure goes to infinity. Considering the problem of state estimation of DES modeled by Petri nets, several contributions have been proposed. In \cite{PNSE1}, the problem of marking estimation based on event observation is discussed, assuming that the net structure is known, and the transition firings can be precisely observed while the initial marking is totally or partially unknown. This work has been extended in \cite{PNSE2} to deal with systems with silent transitions i.e., some transitions are labeled with the empty string. Wang et al. \cite{PNSE4} proposed an approach to refine the marking estimation provided by the method in \cite{PNSE2} by including time information. In \cite{PNSE3}, Bonhomme proposed a method for marking estimation in a timed setting, so that based on a sequence of observation and their firing time, Bonhomme proposed a procedure that determines the set of markings corresponding to the considered observations.

All the works presented above advocate for model-based approaches to deal with DES state estimation. However, even though models offer a clear and formal representation of the system's dynamic, they also encounter notable challenges, especially regarding the complexity and the flexibility when dealing with complex systems. These problems are due to the fact that the number of states of a discrete event model grows exponentially with the number of components, as well as the identification of these models from samples of their languages is also a combinatorial problem of high complexity \cite{DESComp}, and this complexity increases when including timing aspects. In addition, models require a deep understanding of the system dynamic. Therefore, machine learning based approaches are highly demanded, especially those exploiting deep learning tools. Due to their efficient algorithms, and their capacity to deal with heterogeneous data, they can handle highly complex systems where traditional models may be infeasible. In addition, they can also adapt to changes in the system or its environment through continuous training. For this purpose, in this paper, Feed-forward Neural Networks (FNN) are used to deal with Timed Probabilistic Discrete Event Systems (TPDES) state estimation. Based on a data set that contains the functioning history of the system, the proposed model learns to estimate the state of the system directly from this data set without the need for a formal model. Two main cases are considered. \textbf{Case 1:} state estimation of timed probabilistic discrete event systems over observations: in this case, the state estimator computes the probability of being in each state for each new observation. \textbf{Case 2:} state estimation of timed probabilistic discrete event systems over time: in this case, the state estimator provides the probability of being in each state at each clock time increment. These approaches can be viewed as a kind of process discovery (a type of process mining\footnote{Process mining aims to discover, monitor, and improve real processes by extracting
knowledge from event logs readily available in modern information system \cite{processmining}.} that aims to discover real processes merely based on example behaviors stored in event logs \cite{processmining}). So that, the FNN tries to discover the model of the system and understand the system's dynamic purely from raw data, without prior knowledge of the system. To the best of the authors' knowledge, in the literature only few works that use machine learning for the benefit of DES have been proposed. We highlight \cite{SADDEM} that uses recurrent neural networks for online diagnosis of automated production systems of DES class, \cite{Reinforcement1} and \cite{Reinforcement2} that combines reinforcement learning\footnote{Reinforcement learning is a kind of machine learning where the model learns to take decisions by interacting with an environment in order to achieve a specific goal. The agent takes actions, and receives rewards or penalties. The goal is to learn a strategy that maximizes the cumulative reward over time \cite{ML1}, \cite{ML2}.} and Petri nets to curry out the problem of scheduling in manufacturing systems, and \cite{Reachability} and \cite{liveness} that use some machine learning methods for, respectively, probabilistic reachability prediction, and to make liveness decisions for unbounded Petri nets. However, the present paper is the first paper that addresses the problem of TPDES state estimation using neural networks.

The rest of this paper is organised as follows: Section II recalls preliminary notions of TPDES and artificial neural networks. Section III is devoted to the problem statement. Sections IV and V outline the proposed methods for TPDES state estimation. Section VI concludes the paper and discusses future directions. In order to well illustrate the relevance of our approaches, appropriate examples are presented in Sections IV and V.

\section{Preliminaries}
In this section, preliminary notions regarding timed probabilistic discrete event systems and artificial neural networks are presented.
\subsection{Timed Probabilistic Discrete Event Systems}

\begin{definition}
 (Timed Probabilistic Discrete Event Systems) A Timed Probabilistic Discrete Event System (TPDES) is a dynamic system, where the state space is a discrete set and the state changes only at a certain point in time. A TPDES ($\mathcal{S}$, $E$, $\mathcal{O}$, $Obs$, $\mathcal{F}$) is characterized by time semantics that handle the timing aspects, and the following components:
 $\mathcal{S}$ the set of states, E the set of events (or alphabet), $\mathcal{O}$ is the alphabet of observable labels, each label q being generated by the occurrence of a given event $e \in E$ according to a labeling function $Obs: E \rightarrow  \mathcal{O}_{\varepsilon}$, where $\mathcal{O}_{\varepsilon}=\mathcal{O} \cup \{\varepsilon\}$ and $\varepsilon$ is the symbol used to notify that an event is silent i.e., generates no label, $\mathcal{F}$ is a set of probability density functions, each function $\textit{f} \in \mathcal{F}$ specifying the occurrence times of a given event $e \in E$.
\end{definition}

Given an events set $E$, $E^{*}$ is the set of all words (or strings, or sequences of events) on $E$. Let $\sigma \in E^{*}$ be a sequence of events, the length of $\sigma$ is denoted by $|\sigma|$. We note by $(s_{i},e,s_{j})$, where $s_{i}, s_{j} \in \mathcal{S}$ and $e\in E$, the transition from the state $s_{i}$ to the state $s_{j}$ triggered by the event $e$.
Let consider that only some events of $E$ can be observed and the other being silent. In this case, $E=E_{o} \cup E_{u}$, where $E_{o}$ is the set of observable events and $E_{u}$ is the set of unobservable ones. The natural projection $\mathcal{P}: (E \times \mathbb{R}^{+})^{*} \rightarrow (\mathcal{O} \times \mathbb{R}^{+})^{*}$ is used to map the timed sequence of events generated by the system $\sigma^{t}=(e(1),t(1)) (e(2),t(2)) \dots $ to the timed sequence of observations $\nu^{t}=(q(1),t'(1)) (q(2),t'(2)) \dots $. For $\sigma^{t} \in (E \times \mathbb{R}^{+})^{*}$ and $(e,t) \in (E \times \mathbb{R}^{+})$, $\mathcal{P}$ is defined by:

\begin{equation*}
    \begin{cases}
      \mathcal{P}((e,t))=\varepsilon, \hspace{0.5 mm}if \hspace{1 mm}  e \in E_{u};\\
      \mathcal{P}((e,t))=(Obs(e),t), \hspace{0.5 mm}if \hspace{1 mm}  e \in E_{o};\\
      \mathcal{P}(\sigma^{t} (e,t))=\mathcal{P}(\sigma^{t})\mathcal{P}((e,t))
    \end{cases}
\end{equation*}
where $\varepsilon$ represents the empty trace.

In this paper, we consider that we dispose of an external clock $\mathcal{C}$ that measures and tracks the occurrence time of events. This clock is reset after the occurrence of each observable event, i.e., by the occurrence of an observable event, the clock $\mathcal{C}$ is set to zero.

\begin{definition}
   (Timed Run) Consider a timed probabilistic discrete event system ($\mathcal{S}$, $E$, $\mathcal{O}$, $Obs$, $\mathcal{F}$), a timed run $\varrho$ is a sequence of $k+1$ states $s{(i)}\in \mathcal{S}$ and $k$ pairs $(e{(i)}, t{(i)}) \in E \times \mathbb{R}^{+}$, expressed as:
    \begin{equation*}
        \varrho: s{(0)} \xrightarrow{(e{(1)}, t{(1)})} s{(1)} \xrightarrow{(e{(2)}, t{(2)})} \dots s{(k-1)}\xrightarrow{(e{(k)}, t{(k)})}s{(k)}
    \end{equation*}
where $e{(i)}, t{(i)}$, and $ s{(i)}$ refer respectively to the $i^{th}$ event, its occurrence time according to $\mathcal{C}$ and the $i^{th}$ state in $\varrho$. Given a timed run $\varrho$, we define $\sigma(\varrho)$, $\sigma^{t}(\varrho)$, and $\nu^{t}(\varrho)$ as the logical sequence of events, the timed sequence of events, and the timed sequence of observations, respectively, generated during $\varrho$.
\end{definition}

\begin{remark}
A TPDES is characterized by the (usually infinite) set of timed runs it can generate. In the literature several models have been proposed to describe a TPDES with a finite structure. These models include, among others, Markov processes \cite{introdes}, stochastic Petri nets \cite{PNSE1}, \cite{PN33}. The approach presented in this paper for state estimation is based on FNN and can be applied regardless of the model. However, we will use as a way to evaluate the performance of the proposed approach, a particular model called Labeled Timed Probabilistic Automata\footnote{A \textit{Labeled Timed Probabilistic Automata $(LTPA)$} is an extension of the standard finite automata, where each transition ($s_{i}, e, s_{j}$) is endowed with a transition rate $\mu_{i,j}$, so that the $jump$ from the state $s_i$ to the state $s_j$ is triggered by the event $e$ that occurs after a random duration $T$ that is exponentially distributed with the transition rate $\mu_{i,j}$, counted after the system enters to the state $s_i$.}
(LTPA). The choice of this model stems out from the fact that the state probabilities for LTPA can be computed analytically \cite{dim3}.
\end{remark}

\begin{example}
\label{varrhofully}
    Figure \ref{fig:TDES} illustrates two examples of TPDES, modeled with LTPA, where both systems have the same set of states $\mathcal{S}=\{s_{1},s_{2},s_{3},s_{4}\}$, $s_1$ being the initial state, the same set of events $E=\{e_{1},e_{2}, \dots ,e_{10}\}$, the same set of labels $\mathcal{O}=\{a,b,c\}$, and the same set of probability density functions $\mathcal{F}$. In this example, we have considered exponential probability density functions for which parameters are reported on the arcs. For example $e_{1}:a:3$ means that the event $e_{1}$ occurs with a rate equal to 3 and generates the label $a$. The only difference between the two systems is that the system in Figure \ref{PO} disposes on some silent events (events $e_{2}$, $e_{4}$, and $e_{9}$), contrary to the system in Figure \ref{FullyObs}, where all events are observable. For these systems, both the model and the dataset\footnote{Due to the lack of public datasets of TPDES, the data sets that are used in this paper are got by simulating several times the model of the system.} are on our disposal. Both of them are used for illustration and comparison purposes.

\begin{figure}[h!]
\centering
\begin{subfigure}{0.45\textwidth} 
    \centering
    \includegraphics[height=5cm, width=5cm]{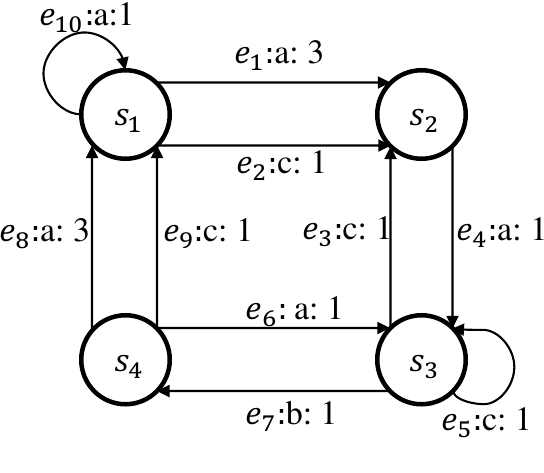}
    \caption{System 1}
    \label{FullyObs}
\end{subfigure}
\hspace{0.05\textwidth} % Adjust space between subfigures
\begin{subfigure}{0.45\textwidth}
    \centering
    \includegraphics[height=5cm, width=5cm]{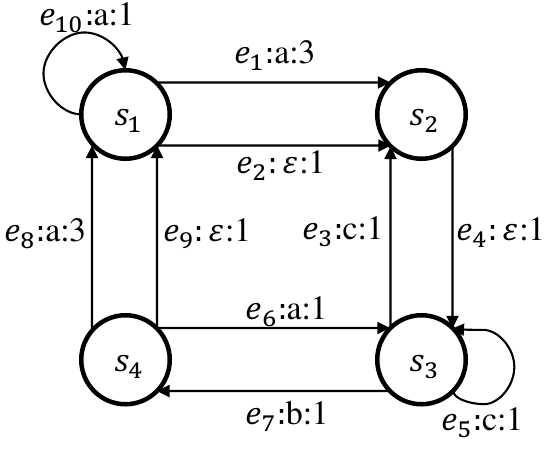}
    \caption{System 2}
    \label{PO}
\end{subfigure}
\caption{Two examples of TPDES modeled by LTPA.}
\label{fig:TDES}
\end{figure}
 As an example, $\varrho_{1}$ is a timed run of system 1 (Figure \ref{FullyObs}) and $\varrho_{1}^{\varepsilon}$ is a timed run of system 2 (Figure \ref{PO}):
    \begin{equation*}
            \varrho_{1}: s_{1} \xrightarrow{(e_1,0.4747)} s_{2} \xrightarrow{(e_4,  0.155)} s_{3} \xrightarrow{(e_7, 1.1232)} s_{4} \xrightarrow{(e_6, 0.3627)} s_{3} \xrightarrow{(e_3,  2.56)} s_{2} \xrightarrow{(e_4,  0.0978)} s_{3}
         \end{equation*}  
\begin{equation*}
           \varrho_{1}^{\varepsilon}: s_{1} \xrightarrow{(e_1,0.1)} s_{2} \xrightarrow{(e_4,  0.3)} s_{3} \xrightarrow{(e_3, 0.5)} s_{2} \xrightarrow{(e_4, 0.3)} s_{3}\xrightarrow{(e_7, 0.4)} s_{4} \xrightarrow{(e_8, 0.1)} s_{1}
         \end{equation*} 
         \begin{equation*}
         \xrightarrow{(e_{10}, 0.2)} s_{1} 
         \end{equation*}
Then, it is:
\begin{equation*}
    \sigma^{t}(\varrho_{1})=(e_1,0.4747) (e_4,  0.155)(e_7, 1.1232)(e_6, 0.3627)(e_3,  2.56) (e_4,  0.0978)
\end{equation*}
\begin{equation*}
    \sigma^{t}(\varrho_{1}^{\varepsilon})=(e_1,0.1)(e_4,  0.3)(e_3, 0.5)(e_4, 0.3)(e_7, 0.4)(e_8, 0.1)(e_{10}, 0.2)
\end{equation*}
\begin{equation*}
    \nu^{t}(\varrho_{1})=(a,0.4747) (a,  0.155)(b, 1.1232)(a, 0.3627)(c,  2.56) (a,  0.0978)
\end{equation*}
\begin{equation*}
    \nu^{t}(\varrho_{1}^{\varepsilon})=(a,0.1)(c, 0.5)(b, 0.4)(a, 0.1)(a, 0.2)
\end{equation*}
\end{example}

\subsection{Neural Networks}
Neural networks are computer models mainly inspired by the connectivity of neuronal cells in the brain. They represent a foundational element of artificial intelligence and machine learning. The main types of neural networks include feed-forward neural networks (FNN), recurrent neural networks (RNN), convolutional neural networks (CNN), and their extensions \cite{ML1}, \cite{ML2}. Neural networks learn to perform tasks through different types of learning such as supervised learning\footnote{In supervised learning, the neural network is trained on a labeled dataset, i.e., each input in the data set is paired with its corresponding output label \cite{ML1}.} and unsupervised learning\footnote{In unsupervised learning, the neural network is trained on data without labels \cite{ML1}.}. In this paper, a feed-forward neural network is trained on a labeled data set. 

\begin{figure}[h]

    \centering
    \includegraphics[height=5.5 cm, width=6 cm]{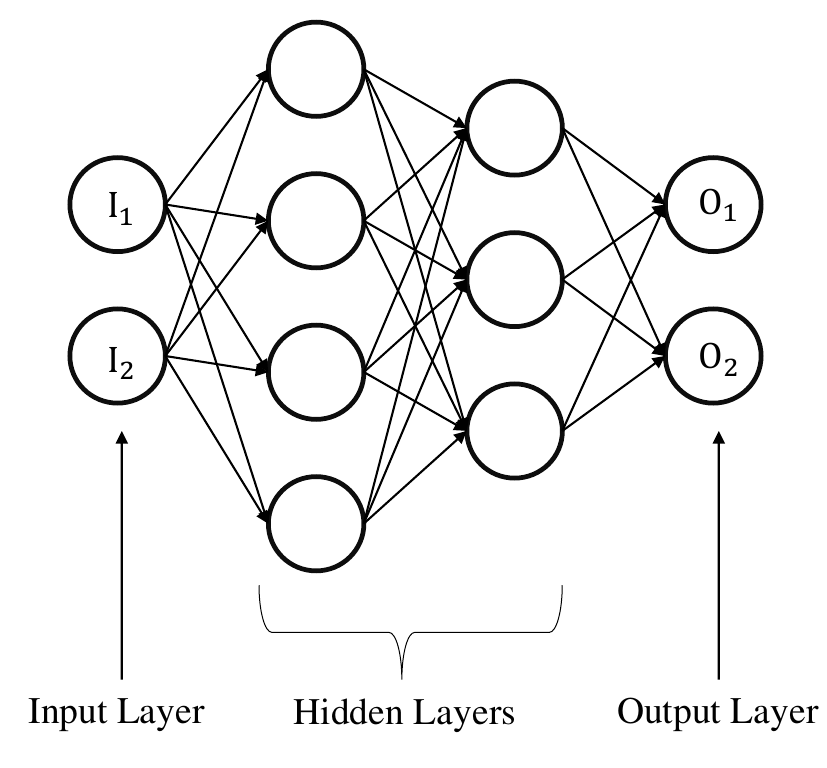}
    \caption{A simple Neural Network.}
    \label{NN}
\end{figure}

FNN are composed of layers (an input layer, one or more hidden layers, and an output layer), where each layer is composed with a specific number of nodes. Each node is linked to the nodes in the preceding and following layer, and each connection has an associated weight and bias. Information propagate from the input layer to the output one. More specificaly, each neuron in an FNN receives inputs from neurons in the preceding layer or from external sources, processes these inputs using an activation function, and then provides outputs to the neurones in the following layer \cite{ML1}. Figure \ref{NN} illustrates an example of a simple FNN with two inputs ($\text{\textit{I}}_{1}$ and $\text{\textit{I}}_{2}$, i.e., the inputs vector is $I=[\textit{I}_{1}$ $ \textit{I}_{2}]^T$), two outputs ($\text{\textit{O}}_{1}$ and $\text{\textit{O}}_{2}$, i.e., the outputs vector is $O=[\textit{O}_{1}$ $ \textit{O}_{2}]^T$), and two hidden layers with four and three neurons respectively. 

The process of developing an FNN model involves three main steps:

\begin{itemize}
    \item \textbf{Training:} The main goal of this step is to allow the FNN to extract relationships between the inputs and their corresponding outputs from the data in the training set \cite{ML1}. For this purpose, at each epoch\footnote{An epoch refers to one pass across the complete training and validation dataset.} the parameters of the FNN (weights and biases) are adjusted (one or several times at each epoch) based on the difference between the actual outputs of the FNN and the target outputs (or the true outputs), in order to minimize the difference between the actual outputs of the FNN and the target outputs. 
    \item \textbf{Validation:} This step is performed in parallel with the training step. It allows to evaluate the performance of the FNN on unseen data during the training. It serves mainly for hyper-parameters (number of layers, number of neurons at each layer, number of epochs, etc) tuning to get a FNN with the best performance, as well as for overfitting\footnote{Overfitting is the case when the model fits too closely or even exactly to its training data engendering negative impacts on its performance on unseen data.} detection \cite{ML1}.
    \item  \textbf{Testing:} This phase allows assessing the performance of the FNN after training and validation. It helps to check how well the FNN generalizes to an unseen dataset that was not used during the training or validation. This step validates the FNN's capability to perform according to the operator's expectations \cite{ML1}.
\end{itemize}

For more information regarding neural networks, the readers are addressed to \cite{ML1}, \cite{ML2}.

\section{Problem Statement}
\label{prblemstatement}
In this work, it is considered that no formal model of the system is available, but the set of states, the set of events, and the set of lables are already known. In addition, we dispose of a dataset containing the functioning history of the system (a detailed description of the dataset will be presented further in this paper). This dataset should be exploited to develop a current state estimator based on FNN. This state estimator is supposed to use both logical and temporal information gathered during the functioning of the system as inputs and provides as output, a probability vector $O \in [0, 1]^{|\mathcal{S}| \times 1}$, where $O_{i}$ represents the probability of the system  being currently in state $s_{i}$. 

\subsection{Assumptions}
In this paper, the following assumptions are considered:

\begin{itemize}
    \item \textbf{$A_{1}:$} The observable language of the system is live.
    \item \textbf{$A_{2}:$} The current state of the system is known during the collection of timed runs and is contained in the raw data used for the FNN elaboration. When a state transition is triggered by an event, either observable or silent, event and its occurrence time are recorded.

    \item \textbf{$A_{3}:$} The dataset used for the FNN elaboration includes sufficient timed runs to accurately and precisely describe the system's behaviour,

    \item \textbf{$A_{4}:$} A detection strategy is implemented to handle out-of-distribution data\footnote{Out-of-distribution data refers to data that differ significantly from the data used to train the FNN \cite{OOD}. In the case of TPDES, an out-of-distribution data may include for example an event that doesn't exist in the dataset, etc.}, i.e., when such out-of-distribution input is identified, it is not forwarded to the FNN. This strategy helps to mitigate the risk of getting erroneous state estimation. 
\end{itemize} 

\subsection{Cases studied, model development and implementation}

In this work, two main cases of state estimation are studied: 
\begin{itemize}
    \item \textbf{Case 1:} State estimation over observations. In this case, assumption \textbf{$A_{5}$} is added to the ones presented previously. \textbf{$A_{5}:$} It is assumed that all events are observable (even two or more events may generate the same output). This assumption is necessary, because, when using FNN, states that are reached exclusively through unobservable events cannot be estimated (For the readers convenience, the assumption \textbf{$A_{5}$} concerns only case 1, case 2 is not concerned.).
    \item \textbf{Case 2:} State estimation over time. In this case, assumption \textbf{$A_{6}$} is considered in addition to $A_1$, $A_2$, $A_3$, and $A_4$. \textbf{$A_{6}:$} It is assumed that the clock advances based on a defined time increment $TI$, and the occurrences of events in the system coincide with multiples of $TI$. This assumption is necessary to update the state estimate so that the state estimation is performed at each clock tick. 
 
\end{itemize}

\begin{figure}[t!]

\begin{center}

    \includegraphics[scale=0.42]{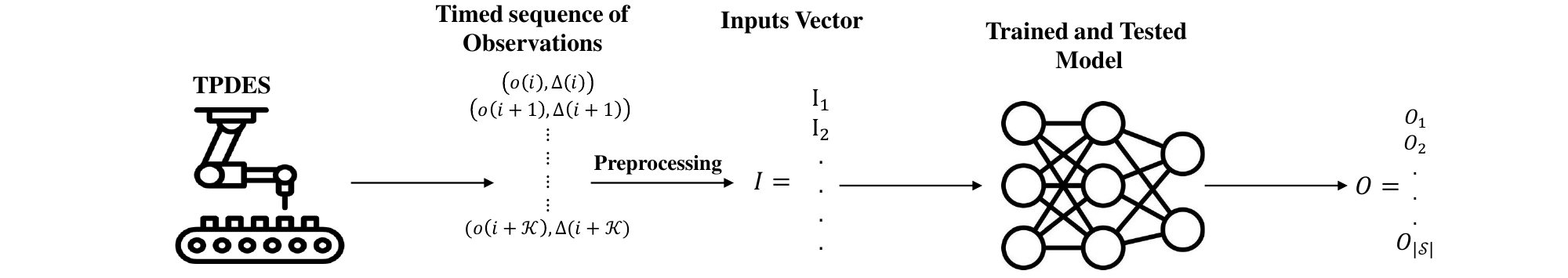}
    \caption{Model implementation}
    \label{MIMPL}
\end{center}
        
\end{figure}

In both cases, the development and implementation steps of the FNN are similar, however, the differences lie in the raw data and the data preprocessing. The proposed approach regarding the deployment of FNN for TPDES state estimation is structured in two main steps:
\begin{itemize}
    \item \textbf{Step 1} (Model development): Timed runs are recorded during multiple runs of the system to collect a raw dataset. This dataset is then preprocessed depending on the proposed approach. After preprocessing, the data is split into training, validation, and testing datasets. Based on these datasets, an appropriate model can subsequently be constructed (Figure \ref{MTV}).

    \begin{figure}[t!]

\begin{center}
    
    \includegraphics[height=7 cm, width=\textwidth]{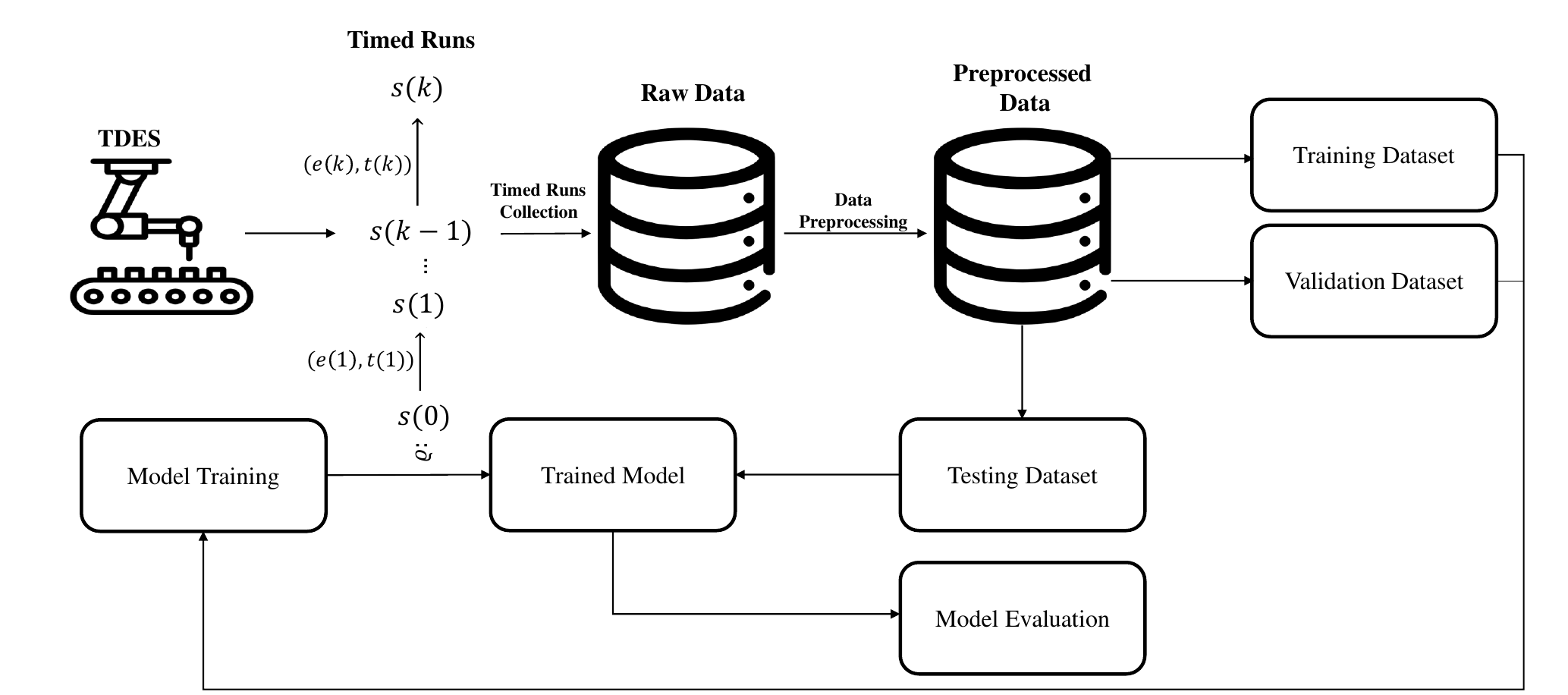}
    \caption{Model development}
\label{MTV}
\end{center}
    
\end{figure}

    \item \textbf{Step 2} (Model implementation): In this step, the resulting FNN from step 1 is implemented to estimate the current state of the system. For this purpose, an input vector $I=[I_{1}, I_{2}, \dots]^{T}$ is constructed at each new observation if the state estimate is performed over observations, or at each new clock tick if the state estimate is performed over time based on the logical and timed information collected during the system's operation (this vector will be detailed further in this paper). The setup of $I$ mirrors the preprocessing methods used for the raw data in the data preprocessing step. Then, $I$ is fed into the FNN. The FNN then treats this vector and outputs $O$ (Figure \ref{MIMPL}).
\end{itemize}

\section{State Estimation of TPDES over Observations}
In this section, the proposed approach of TPDES state estimation over observations is presented.
\subsection{Data Description and Data Preprocessing}

In our approach, a dataset composed of timed runs is used. Let denote by $\mathcal{RD}$ the set of raw data, where $\mathcal{RD}=\{\varrho_1, \dots \varrho_{rd}\}$, so that $rd=|\mathcal{RD}|$ represents the total number of samples within $\mathcal{RD}$. Each sample in $\mathcal{RD}$ is a timed run of the system. These timed runs are collected during different scenarios, which we assume represent a significant sample of the system’s behavior.

Formatting data is mandatory to make them exploitable for the FNN. Algorithm~\ref{Algofullyobs} is proposed for this purpose. This algorithm takes as inputs $\mathcal{RD}$ and a hyper-parameter $\mathcal{K} \in \mathbb{N}^{*}$\footnote{$\mathbb{N}^{*}$ is the set of non-null natural numbers.} that will be detailed in the next paragraph, and provides as outputs DataInputs=$[V_{I_1},V_{I_2},  \dots, V_{I_i}, \dots ]$ and DataOutputs=$[V_{O_1},V_{O_2}, \dots, V_{O_i}, \dots ]$ that represent respectively the input and output vectors that will be used for the FNN development, where $V_{I_i}$ is an input vector and $V_{O_i}$ is its corresponding output vector. This algorithm converts each timed run $\varrho_{i}$ into $|\varrho_{i}|+1$ samples. The hyper-parameter $\mathcal{K}$ is tuned by tests to find out the value that provides the desired performance.

This algorithm operates as follows: initially, for each timed run $\varrho_{i}$, the first vector that is constructed is an all zero vector, and its corresponding output vector is generated using a function called $Class$: this function takes $s{(0)}$, which represents the initial state in the timed run, and transform it to a one column vector of size $|\mathcal{S}|$, so that all the elements of this vector are zeros except for the element at the index corresponding to the state $s{(0)}$, which is set to one. Consider the system in Figure \ref{FullyObs} as an example, \textit{Class}($s_{1})$=$[1$ $0$ $0$ $0]^{T}$, \textit{Class}($s_{2})$=$[0$ $1$ $0$ $0]^{T}$, \textit{Class}($s_{3})$=$[0$ $0$ $1$ $0]^{T}$, and \textit{Class}($s_{4})$=$[0$ $0$ $0$ $1]^{T}$. The purpose of this vector is to enable the FNN to estimate the initial state of the system so that when it is implemented and before the occurrence of any event, an all zero vector will be fed to the FNN and the output vector will be the initial state probability. Further, a loop iterates over all observations generated during $\varrho_{i}$. For each observation $q{(i)}$, the algorithm takes the observation $q{(i)}$ and its occurrence time $t{(i)}$ and the $\mathcal{K}-1$ previous observations and their occurrence time and form an input vector $V_{I_{i}}=[q{(i-\mathcal{K}+1)}$ $t{(i-\mathcal{K}+1)} \dots q{(i)}$ $ t{(i)}]^{T}$. For $i=1, \dots, \mathcal{K}-1$, the vector $V_{I_i}$ is completed with zero entries. The corresponding output vector $V_{O_{i}}$ of $V_{I_{i}}$ is $V_{O_{i}}=Class(s{(i)})$, where $s{(i)}$ is the state reached that far in the timed run. Then, these operations are repeated for all the timed runs in $\mathcal{RD}$ and the algorithm returns DataInputs and DataOutputs. 

Based on this dataset the FNN will extract two main features of the system: the order of the observations and the temporal constraints between them and recognizes the relation between these features and the states of the system, in such a way that the FNN will estimate the states of the system based on the order of the observations and the time between consecutive observations. Analyzing both information allows the FNN to understand the system's dynamic, which will give it the ability to estimate effectively the current state of the system after implementation.

\begin{algorithm}[H]

\textbf{Inputs:} $\mathcal{RD}=\{\varrho_1, \dots \varrho_{rd}\}$, $\mathcal{K}$\\
\textbf{Outputs:} DataInputs, DataOutputs\\
DataInputs $\leftarrow [$ $]$, DataOutputs $\leftarrow [$ $]$, $V_{I} \leftarrow [$ $] $, $V_{O} \leftarrow [$ $]$, $rd\leftarrow|\mathcal{RD}|$\\
\For{i=1 to rd}{
$V_{I} \leftarrow $[$0_{1 \times (2 \times \mathcal{K})}]^{T}$, $V_{O} \leftarrow $ \textit{Class}($s{(0)}$)\\
DataInputs$\leftarrow$DataInputs $\cup$ $V_{I}$, DataOutputs$\leftarrow$DataOutputs $\cup$ $V_{O}$\\
$k\leftarrow|\sigma(\varrho_{i})|$\\
\For{j=1 to k}{
\eIf{j $< \mathcal{K}$ }{
$V_{I}\leftarrow[q{(1)}$ $t{(1)}$ $\dots$ $q{(j)}$ $t{(j)}]^{T}$\\
$V_{I}\leftarrow[0_{1 \times (2 \times \mathcal{K} - |V_{I}|)}$ $ V_{I}]^{T}$\\
$V_{O}\leftarrow$\textit{Class}($s{(j)}$)
}{
$V_{I}\leftarrow[q{(j-\mathcal{K}+1)}$ $t{(j-\mathcal{K}+1)}$ $\dots$ $q{(j)}$ $t{(j)}]^{T}$\\
$V_{O}\leftarrow$\textit{Class}($s{(j)}$)
}
DataInputs$\leftarrow$DataInputs $\cup$ $V_{I}$\\
DataOutputs$\leftarrow$DataOutputs $\cup$ $V_{O}$
}
}

\caption{Data formatting algorithm for TPDES state estimation over observations.}
\label{Algofullyobs}
\end{algorithm}

\begin{example}

    Let consider the system modeled in Figure \ref{FullyObs}, and let us take the run $\varrho_{1}$ as an example. Let $\mathcal{K}=3$. By applying Algorithm \ref{Algofullyobs} to $\varrho_{1}$, the resulting DataInputs and DataOutputs are as follows:

DataInputs =
$$\left[
\begin{array}{c c c c c c c c c}
V_{I_1}&V_{I_2}&V_{I_3}&V_{I_4}&V_{I_5}&V_{I_6}&V_{I_7}\\
\downarrow &\downarrow&\downarrow&\downarrow&\downarrow&\downarrow&\downarrow\\

\begin{bmatrix}
0\\ 0\\ 0\\ 0\\ 0\\ 0
\end{bmatrix}&
\begin{bmatrix}
0\\ 0\\ 0\\ 0\\ a\\ 0.4747
\end{bmatrix}&
\begin{bmatrix}
0\\ 0\\ a\\ 0.4747\\ a\\ 0.155
\end{bmatrix}&

\begin{bmatrix}
a\\ 0.4747\\ a\\ 0.155\\b\\1.1232
\end{bmatrix}&
\begin{bmatrix}
 a\\ 0.155\\b\\1.1232\\a\\0.3627
\end{bmatrix}&
\begin{bmatrix}
b\\1.1232\\a\\0.3627\\c\\2.56
\end{bmatrix}&
\begin{bmatrix}
a\\0.3627\\c\\2.56\\a\\0.0978
\end{bmatrix}
\end{array} \right],$$

$$ DataOutputs = \left[ \begin{array}{c c c c c c c c c}

V_{O_1}&V_{O_2}&V_{O_3}&V_{O_4}&V_{O_5}&V_{O_6}&V_{O_7}\\
\downarrow &\downarrow&\downarrow&\downarrow&\downarrow&\downarrow&\downarrow\\

\begin{bmatrix}
1\\0\\0\\0
\end{bmatrix}&
\begin{bmatrix}
0\\1\\0\\0
\end{bmatrix}&
\begin{bmatrix}
0\\0\\1\\0
\end{bmatrix}&
\begin{bmatrix}
0\\0\\0\\1
\end{bmatrix}&
\begin{bmatrix}
0\\0\\1\\0
\end{bmatrix}&
\begin{bmatrix}
0\\1\\0\\0
\end{bmatrix}&
\begin{bmatrix}
0\\0\\1\\0
\end{bmatrix}

\end{array} \right].$$
\end{example}

It is worth noting that $a$, $b$, $c$ are encoded as numerical data and not categorical ones.

\subsection{Model Building and Implementation}
After the completion of the data preprocessing step, the next phase is the model building. For this purpose, DataInputs and DataOutputs are devised to three main sets: training set, validation set and testing set. A rigorous step of hyper-parameter tuning is mandatory to find out the best FNN in order to ensure optimal performance. After all the steps have been completed successfully, the resulting FNN, with input vector of size $2 \times \mathcal{K}$ and output vector of size $|\mathcal{S}|$ is the probabilistic state estimator developed for TPDES state estimation over observations.

Then, the resulting FNN is implemented, and following each observation $q$, an input vector $I$ is constructed, where $I$ contains the current observation, its occurrence time and the $\mathcal{K}-1$ last observations and their occurrence time ($I$ is constructed with the same way used to construct $V_I$).  
This vector is fed to the FNN, which will provide $O$ as output.

\begin{example}
\label{example fulyy}
Let us consider the system modeled in Figure \ref{FullyObs}. The raw data contain 400 timed runs. Various values of the parameter $\mathcal{K}$ are tested, the optimal results were obtained with $\mathcal{K}=3$. After data formatting the obtained data contains $9150$ data points, where $7250$ (300 timed runs) are used for training, $800$ (50 timed runs) for validation and $1100$ (50 timed runs) for testing. The architecture of the FNN used is detailed in Table \ref{resultnn}. The architecture contains three dense layers with 64, 32, and 4 nodes respectively. The activation function ReLU
is used for all Dense layers except the output layer that uses a Softmax. Dropout layer, with a drop out rate of 0.4, is added after the first layer, in order to prevent over fitting.
The training comprises 250 epochs. Adam optimizer, a learning rate of 0.001, categorical crossentropy loss and batch size of 500 are used (For more information about the components used in this FNN, such as the activation functions, the loss, the optimizer,...etc, we encourage the readers to conduct \cite{ML1}, \cite{ML2}, \cite{web1}, and \cite{web2}). The results of the training are depicted in Figure \ref{ACCTRAINING}.
\begin{table}[t!]
\begin{tabular}{c c c c} 
 \hline
 Layer & Type & Activation & Output \\
No. &  & Function & Shape \\
 \hline
 1 & InputLayer&-&6\\ 
 2 & Dense & ReLu & 64 \\ 
 3 & Dropout & - & 64 \\
 4 & Dense & ReLu & 32 \\
 5 & Dense & Softmax & 4 \\
 \hline
\end{tabular}
 \caption{The FNN architecture for TPDES state estimation over observations.} 
    \label{resultnn}
\end{table}

\begin{figure}[h!]
\begin{center}
    \includegraphics[height=5 cm, width=0.7\textwidth]{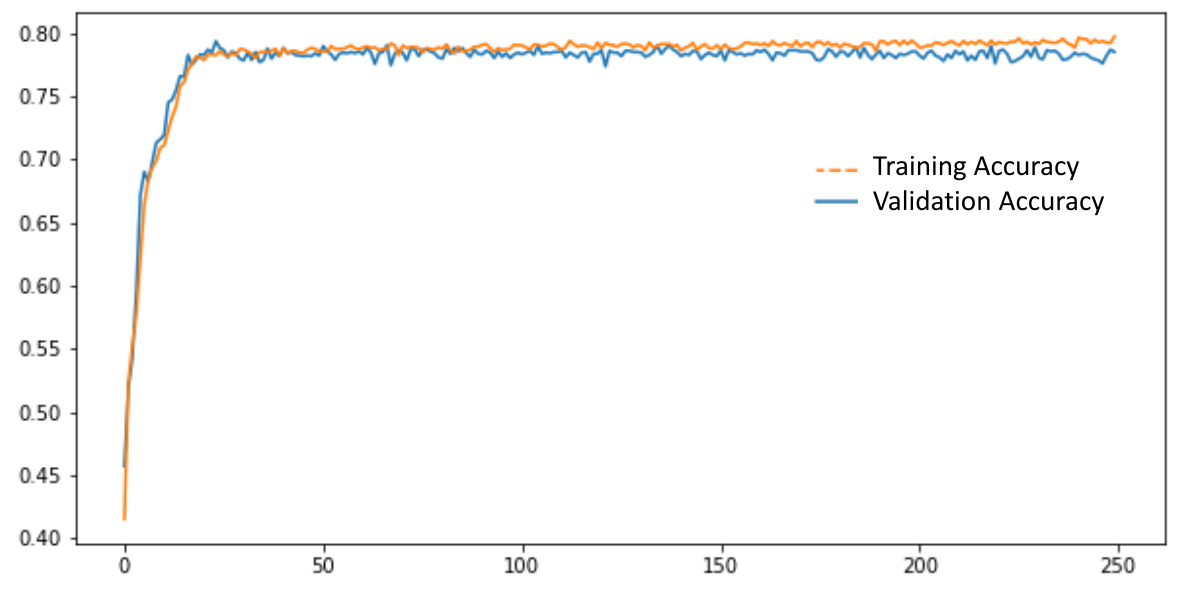}
    \caption{The training and validation accuracy (TPDES state estimation over observations). }
         \label{ACCTRAINING}
\end{center}
   
\end{figure}
The training and validation accuracy reach approximately 0.80 and 0.79, respectively. The next step involves testing the FNN. For evaluating the FNN's performance, we consider the estimation to be accurate if the true state is assigned the highest probability among all possible states, and the accuracy is computed by dividing the number of good estimations by the total number of instances in the testing data set. Consequently, the testing accuracy is almost $80 \%$. For comparison purposes, we compare the state estimation based on the FNN and the model based state estimation approach (MBSE) presented in \cite{jdeds1}, \cite{jdeds2}. Note that by MBSE the state estimate is reconstructed at each instant $t$ contrary to the approach presented in this section that reconstructs the state estimate at each new observation. Therefore and for a fair comparison, we compare the state estimate based on the FNN and the MBSE approach only at the occurrence time of observations and at $t=0$. The comparison is done over all the sequences extracted from the testing dataset, and the Mean Absolute Error $(MAE)$\footnote{The MAE measures the average absolute difference between the state estimations provided after processing the FNN outputs and the MBSE approach. Let $\Pi = [\Pi(s_1), \dots, \Pi(s_{|\mathcal{S}|})]$ and $\hat{\Pi} = [\hat{\Pi}(s_1), \dots, \hat{\Pi}(s_{|\mathcal{S}|})]$ be the probability vectors representing one of the state estimates based on the FNN and MBSE approaches, respectively. The MAE between these two vectors is computed by: $\frac{1}{|\mathcal{S}|} \sum_{i=1}^{|\mathcal{S}|} |\hat{\Pi}(s_i) - \Pi(s_i)|$. Given $m$ state estimates, the overall MAE is calculated by averaging the individual MAEs:
$\frac{1}{m} \sum_{j=1}^{m} MAE_j$.} is computed between the estimations provided by the FNN and the MBSE approach. As much as the MAE is close to 0 as much as the state estimations provided by both approaches are identical. The $MAE$ obtained is almost $1.5 \%$, so, the results obtained using the FNN are quite similar to those derived from the MBSE approach. Based on these results, we can conclude that the network developed exhibits good performance.

As an example, let consider a scenario, where the following timed sequence of observations is recorded:
\begin{align*}
    \nu^{t}=(c,0.387)(a,4.1161)(c,0.099)(c,0.2257)(a, 0.2274)(b,0.0096)(a,0.4297)(c,0.3494)\\
     (a,1.3324)(c,0.033)(a,0.1197)(b,0.1738)(a,0.532)(a,0.1834)(a, 0.1128)
\end{align*}

Applying our network, the results of the state estimation are depicted in Figures \ref{estimate s1 fully}, \ref{estimate s2 fully}, \ref{estimate s3 fully} and \ref{estimate s4 fully} with blue dots (the results of the state estimation using the MBSE approach are also reported in orange dashed curve). 

\begin{figure}[h!]

      \centering
      \includegraphics[scale=0.7]{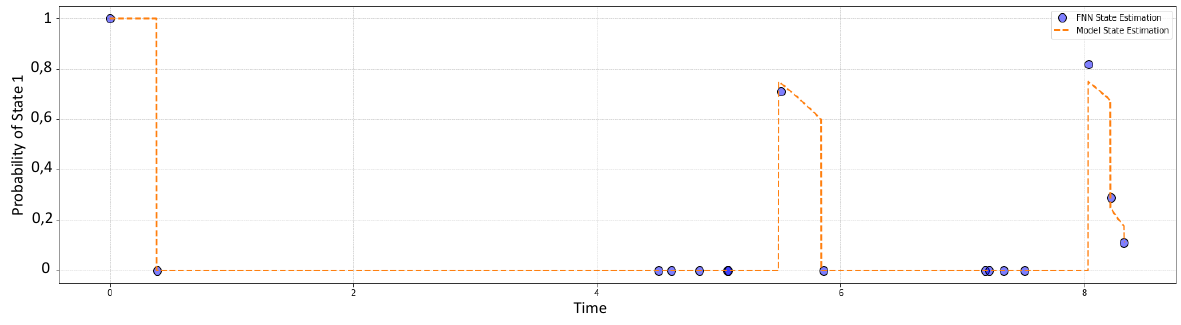}
      \caption{The probability of state $s_1$}
      \label{estimate s1 fully}
\end{figure}
\begin{figure}[h!]
      \centering
      \includegraphics[scale=0.7]{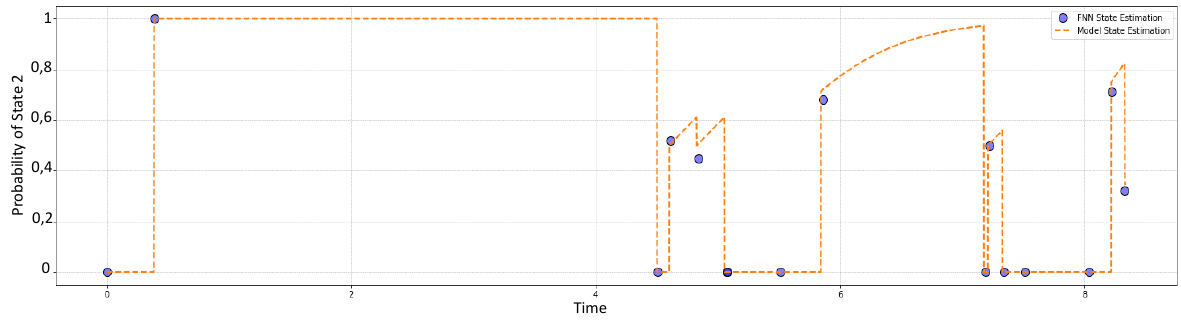}
      \caption{The probability of state $s_2$}
      \label{estimate s2 fully}
\end{figure}
 \begin{figure}[h!]
      \centering
      \includegraphics[scale=0.7]{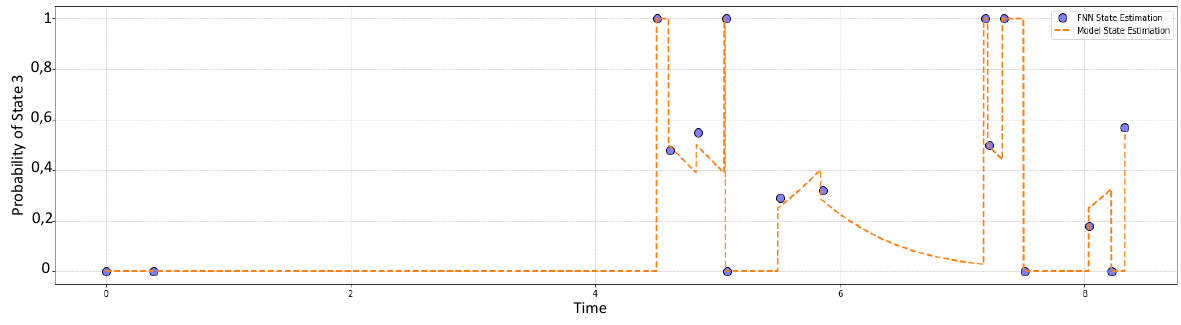}
      \caption{The probability of state $s_3$}
      \label{estimate s3 fully}
\end{figure}
\begin{figure}[h!]
      \centering
      \includegraphics[scale=0.7]{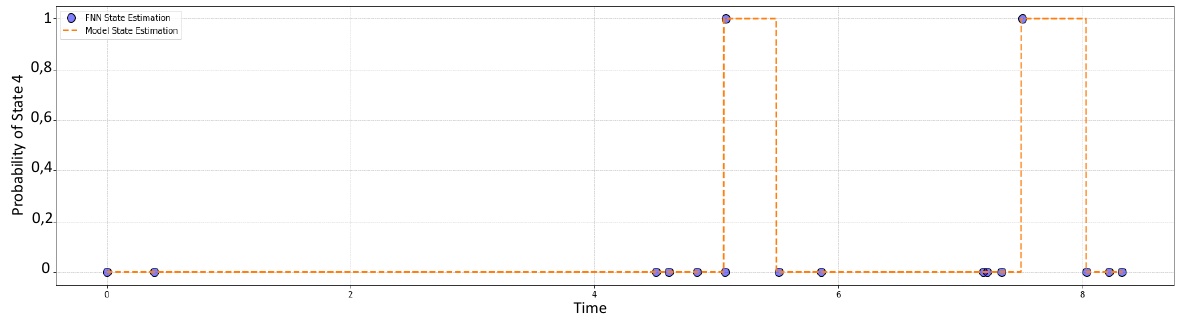}
      \caption{The probability of state $s_4$}
      \label{estimate s4 fully}
\end{figure}

From these figures, we can remark that the results based on the FNN are very close to those provided by the MBSE highlighting the effectiveness and accuracy of the FNN approach presented in this section.
\end{example}

\section{State Estimation of TPDES over Time}
The following section presents the proposed approach of TPDES state estimation over time.
\subsection{Data Description and Data Preprocessing}
Similar to the state estimation of TPDES over observations, the dataset that will be used here consists also of timed runs. Let denote by $\mathcal{RD}^{\varepsilon}$ the set of raw data used for this case, where $\mathcal{RD}^{\varepsilon}=\{\varrho_{1}^{\varepsilon}, \dots \varrho_{rn}^{\varepsilon}\}$, so that $rn=|\mathcal{RD}^{\varepsilon}|$ represents the total number of timed runs contained within $\mathcal{RD}^{\varepsilon}$.

Algorithm \ref{Algopartilyobs} is used for data preprocessing. The inputs of this algorithms are $\mathcal{RD}^{\varepsilon}$, $TI$, and a hyper-parameter $\mathcal{K}^{\varepsilon} \in \mathbb{N}^{*}$. The outputs are DataInputs$^{\varepsilon}$=$[V_{I_1}^{\varepsilon},V_{I_2}^{\varepsilon},  \dots, V_{I_i}^{\varepsilon}, \dots ]$ and DataOutputs$^{\varepsilon}$=$[V_{O_1}^{\varepsilon},V_{O_2}^{\varepsilon},  \dots, V_{O_i}^{\varepsilon}, \dots ]$, so that $V_{I_i}^{\varepsilon}$ is an input vector and $V_{O_i}^{\varepsilon}$ is its corresponding output vector. 

Initially, for each timed run, the first vector constructed is an all-zero vector, along with its corresponding output vector, $Class(s{(0)})$. Next, for each $e{(j)}$, Algorithm \ref{Algopartilyobs} calculates the number of clock ticks that have elapsed since $e{(j-1)}$ (or since 0, for $j=1$). For each tick, the algorithm creates an input vector $V^{\varepsilon}_{I_i}$ that contains the $\mathcal{K}^{\varepsilon}$ last observations that exist before $e{(j)}$, their occurrence time and the elapsed time at this tick since the last observation, along with its corresponding output vector $V^{\varepsilon}_{o_i}=Class(s)$, where $s$ is the current state at this tick. If the size of $V^{\varepsilon}_{I_i}$ is less than $2\times\mathcal{K}^{\varepsilon} + 1$, the vector is padded with zeros. Subsequently, these procedures are replicated for all the timed runs in $\mathcal{RD}^{\varepsilon}$ to generate DataInputs$^{\varepsilon}$ and DataOutputs$^{\varepsilon}$.

In this case, based on this dataset, the network will extract three main features regarding the system's dynamic: the order of the observations, the temporal constraints between them, and the behaviour of the system when it remains silent. Including the elapsed time since the last observation in the input vector provides valuable temporal information, so that, in addition to the order of the observations and the time constraints between consecutive observations, the FNN can recognize the dynamic of the system even when it remains silent.

\begin{algorithm}[H]
\textbf{Inputs:} $\mathcal{RD}^{\varepsilon}=\{\varrho_{1}^{\varepsilon}, \dots \varrho_{rn}^{\varepsilon}\}$, $TI$, $\mathcal{K}^{\varepsilon}$\\
\textbf{Outputs:} DataInputs$^{\varepsilon}$, DataOutputs$^{\varepsilon}$\\
DataInputs$^{\varepsilon}\leftarrow$ $[$ $]$, DataOutputs$^{\varepsilon}\leftarrow [$ $]$, $\mathcal{K}^{\varepsilon}$-last-observations$\leftarrow [$ $]$, $V_{I}^{\varepsilon}\leftarrow [$ $]$, $V_{O}^{\varepsilon}\leftarrow [$ $]$, $rn\leftarrow |\mathcal{RD}^{\varepsilon}|$\\
\For{i=1 to rn}{
$V_{I}^{\varepsilon}\leftarrow$[$0_{1 \times (2 \times \mathcal{K}^{\varepsilon}+1)}$$]^{T}$, $V_{O}^{\varepsilon}\leftarrow$ \textit{Class}($s{(0)}$)\\
DataInputs$^{\varepsilon}\leftarrow$DataInputs$^{\varepsilon}$ $\cup$ $V_{I}^{\varepsilon}$, DataOutputs$^{\varepsilon}\leftarrow$DataOutputs$^{\varepsilon}$ $\cup$ $V_{O}^{\varepsilon}$\\
$k\leftarrow|\sigma(\varrho_{i}^{\varepsilon})|$\\
\For{j=1 to k}{

Compute number-of-ticks\\
\For{nt=1 to number-of-ticks-1}{
\If{$ |\mathcal{K}^{\varepsilon}$\text{-last-observations$|$} $<$ $2 \times \mathcal{K}^{\varepsilon}$ }{ complete $ \mathcal{K}^{\varepsilon}$\text{-last-observations} with zeros until it reaches the dimension $2 \times \mathcal{K}^{\varepsilon}$}
elapsed-time$\leftarrow$elapsed-time + $TI$\\
$V_{I}^{\varepsilon}\leftarrow$[$\mathcal{K}^{\varepsilon}$\text{-last-observations} elapsed-time], $V_{O}^{\varepsilon}\leftarrow Class(s{(j-1)})$ \\
DataInputs$^{\varepsilon}\leftarrow$DataInputs$^{\varepsilon}$ $\cup$ $V_{I}^{\varepsilon}$, DataOutputs$^{\varepsilon}\leftarrow $DataOutputs$^{\varepsilon}$ $\cup$ $V_{O}^{\varepsilon}$
}

\eIf{$Obs(e{(j)}) \neq \varepsilon$}{ 
Remove the two first elements of $\mathcal{K}^{\varepsilon}$\text{-last-observations}\\
$\mathcal{K}^{\varepsilon}$\text{-last-observations}$\leftarrow$[$\mathcal{K}^{\varepsilon}$\text{-last-observations} $Obs(e{(j)})$ $t{(j)}$] \\
elapsed-time$\leftarrow0$
}{
elapsed-time$\leftarrow$elapsed-time+$TI$
}
$V_{I}^{\varepsilon}\leftarrow$[$\mathcal{K}^{\varepsilon}$\text{-last-observations} elapsed-time], $V_{O}^{\varepsilon}\leftarrow Class(s{(j)})$ \\
DataInputs$^{\varepsilon}\leftarrow$DataInputs$^{\varepsilon}$ $\cup$ $V_{I}^{\varepsilon}$, DataOutputs$^{\varepsilon}\leftarrow$DataOutputs$^{\varepsilon}$ $\cup$ $V_{O}^{\varepsilon}$
}
}

\caption{Data formatting algorithm for TPDES state estimation over time.}
\label{Algopartilyobs}
\end{algorithm}

\begin{example}
Let consider the system presented in the Figure \ref{PO}, and take $\varrho_{1}^{\varepsilon}$ as example. By applying Algorithm \ref{Algopartilyobs} with $\mathcal{K}^{\varepsilon}=5$ to $\varrho_{1}^{\varepsilon}$, the resulting DataInputs$^{\varepsilon}$ and DataOutputs$^{\varepsilon}$ are as follows:

$$DataInputs^{\varepsilon} = \left[ \begin{array}{c c c c c c c c c c c c}
V_{I_1}^{\varepsilon}&V_{I_2}^{\varepsilon}&V_{I_3}^{\varepsilon}&V_{I_4}^{\varepsilon}&V_{I_5}^{\varepsilon}&V_{I_6}^{\varepsilon}&V_{I_7}^{\varepsilon}&V_{I_8}^{\varepsilon}&V_{I_9}^{\varepsilon}&V_{I_{10}}^{\varepsilon}&V_{I_{11}}^{\varepsilon}&V_{I_{12}}^{\varepsilon}\\

\downarrow &\downarrow&\downarrow&\downarrow&\downarrow&\downarrow&\downarrow&\downarrow&\downarrow&\downarrow&\downarrow&\downarrow\\
\begin{bmatrix}
0\\ 0\\ 0\\ 0\\ 0\\ 0\\ 0\\ 0\\ 0\\ 0\\ 0
\end{bmatrix}&

\begin{bmatrix}
0\\ 0\\ 0\\ 0\\ 0\\ 0\\ 0\\ 0\\ a\\ 0.1\\ 0
\end{bmatrix}&
\begin{bmatrix}
0\\ 0\\ 0\\ 0\\ 0\\ 0\\ 0\\ 0\\ a\\ 0.1\\ 0.1
\end{bmatrix}&
\begin{bmatrix}
0\\ 0\\ 0\\ 0\\ 0\\ 0\\ 0\\ 0\\ a\\ 0.1\\ 0.2
\end{bmatrix}&
\begin{bmatrix}
0\\ 0\\ 0\\ 0\\ 0\\ 0\\ 0\\ 0\\ a\\ 0.1\\ 0.3
\end{bmatrix}&
\begin{bmatrix}
0\\ 0\\ 0\\ 0\\ 0\\ 0\\ 0\\ 0\\ a\\ 0.1\\ 0.4
\end{bmatrix}&

\begin{bmatrix}
 0\\ 0\\ 0\\ 0\\ 0\\ 0\\ a\\ 0.1 \\ c\\ 0.5 \\0
\end{bmatrix}&
\begin{bmatrix}
 0\\ 0\\ 0\\ 0\\ 0\\ 0\\ a\\ 0.1 \\ c\\ 0.5 \\0.1
\end{bmatrix}&
\begin{bmatrix}
 0\\ 0\\ 0\\ 0\\ 0\\ 0\\ a\\ 0.1 \\ c\\ 0.5 \\0.2
\end{bmatrix}&
\begin{bmatrix}
 0\\ 0\\ 0\\ 0\\ 0\\ 0\\ a\\ 0.1 \\ c\\ 0.5 \\0.3
\end{bmatrix}&
\begin{bmatrix}
 0\\ 0\\ 0\\ 0\\ a\\ 0.1 \\ c\\ 0.5 \\b\\0.4\\0
\end{bmatrix}&
\begin{bmatrix}
 0\\ 0\\ a\\ 0.1 \\ c\\ 0.5 \\b\\0.4\\a \\0.1\\0
\end{bmatrix}
\end{array} \right]$$\\
$$DataOutputs^{\varepsilon} = \left[ \begin{array}{c c c c c c c c c c c c c c c}
V_{O_1}^{\varepsilon}&V_{O_2}^{\varepsilon}&V_{O_3}^{\varepsilon}&V_{O_4}^{\varepsilon}&V_{O_5}^{\varepsilon}&V_{O_6}^{\varepsilon}&V_{O_7}^{\varepsilon}&V_{O_8}^{\varepsilon}&V_{O_9}^{\varepsilon}&V_{O_{10}}^{\varepsilon}&V_{O_{11}}^{\varepsilon}&V_{O_{12}}^{\varepsilon}\\

\downarrow &\downarrow&\downarrow&\downarrow&\downarrow&\downarrow&\downarrow&\downarrow&\downarrow&\downarrow&\downarrow&\downarrow\\
\begin{bmatrix}
1\\ 0 \\0\\0
\end{bmatrix}&
 \begin{bmatrix}
0\\ 1 \\0\\0
\end{bmatrix}&
 \begin{bmatrix}
0\\ 1 \\0\\0
\end{bmatrix}&
 \begin{bmatrix}
0\\ 1 \\0\\0
\end{bmatrix}&
 \begin{bmatrix}
0\\ 0 \\1\\0
\end{bmatrix}&
 \begin{bmatrix}
0\\ 0 \\1\\0
\end{bmatrix}&
 \begin{bmatrix}
0\\ 1 \\0\\0
\end{bmatrix}&
 \begin{bmatrix}
0\\ 1 \\0\\0
\end{bmatrix}&
 \begin{bmatrix}
0\\ 0 \\1\\0
\end{bmatrix}&
 \begin{bmatrix}
0\\ 0 \\1\\0
\end{bmatrix}&
 \begin{bmatrix}
0\\ 0 \\0\\1
\end{bmatrix}&
\begin{bmatrix}
1\\ 0 \\0\\0
\end{bmatrix}&

\end{array} \right]$$

\end{example}

\subsection{Model Building and Implementation}
DataInputs$^{\varepsilon}$ and DataOutputs$^{\varepsilon}$ are separated into three sets for training, validation and testing. A rigorous hyper-parameter tuning should also be done to find out the optimal FNN. After all the steps have been completed successfully, the resulting FNN, with $2 \times \mathcal{K}^{\varepsilon}$ +1 inputs and $|\mathcal{S}|$ outputs, is ready for TPDES state estimation over time.\\
Then, the resulting FNN is implemented, and following each clock tick, an input vector $I$ is constructed, where $I$ contains the $\mathcal{K}^{\varepsilon}$ last observations, their occurrence time and the elapsed time since the last observation according to $\mathcal{C}$.
This vector is fed to the FNN, which will provide $O$ as output.

\begin{example}
\label{example parttt}
Let take the system modeled in Figure \ref{PO}. The raw data contain 250 timed runs. Various values of the parameter $\mathcal{K}^{\varepsilon}$ are tested, the best results were obtained with $\mathcal{K}^{\varepsilon}=5$. After data preprocessing the resulting data contains $18365$ data points, where $15175$ are used for training (200 timed runs), $1562$ for validation (25 timed runs) and $1628$ for testing (25  timed runs). The architecture of the FNN used is detailed in Table \ref{resultnn_par}. The architecture contains six dense layers with 256, 128, 128, 64, 64, and 4 nodes respectively. The activation function ReLU is used for all Dense layers except the output layer that use a Softmax. Dropout layers, with a drop out rate of 0.6 are added after the first, the second and the third layers.
The training comprises 50 epochs. Adam optimizer, a learning rate of 0.001, binary crossentropy loss and batch size of 100 are used. The results of the training are depicted in Figure \ref{ACCTRAINING_par}.
\begin{table}[t!]
\begin{tabular}{c c c c} 
 \hline
 Layer & Type & Activation & Output \\
No. &  & Function & Shape \\
 \hline
 1 & InputLayer&-&9\\ 
 2 & Dense & ReLu & 256 \\ 
 3 & Dropout & - & 256 \\
 4 & Dense & ReLu & 128 \\
 5 & Dropout & - & 128 \\
 6 & Dense & ReLu & 128 \\
 7 & Dropout & - & 128 \\
 8 & Dense & - & 64 \\
 9 & Dense & ReLu & 64 \\
 10 & Dense & Softmax & 4 \\
 \hline
\end{tabular}
 \caption{The neural network architecture for TPDES wstate estimation over Time.} 
    \label{resultnn_par}
\end{table}

\begin{figure}[h!]
\begin{center}
    \includegraphics[height=5 cm, width=0.7\textwidth]{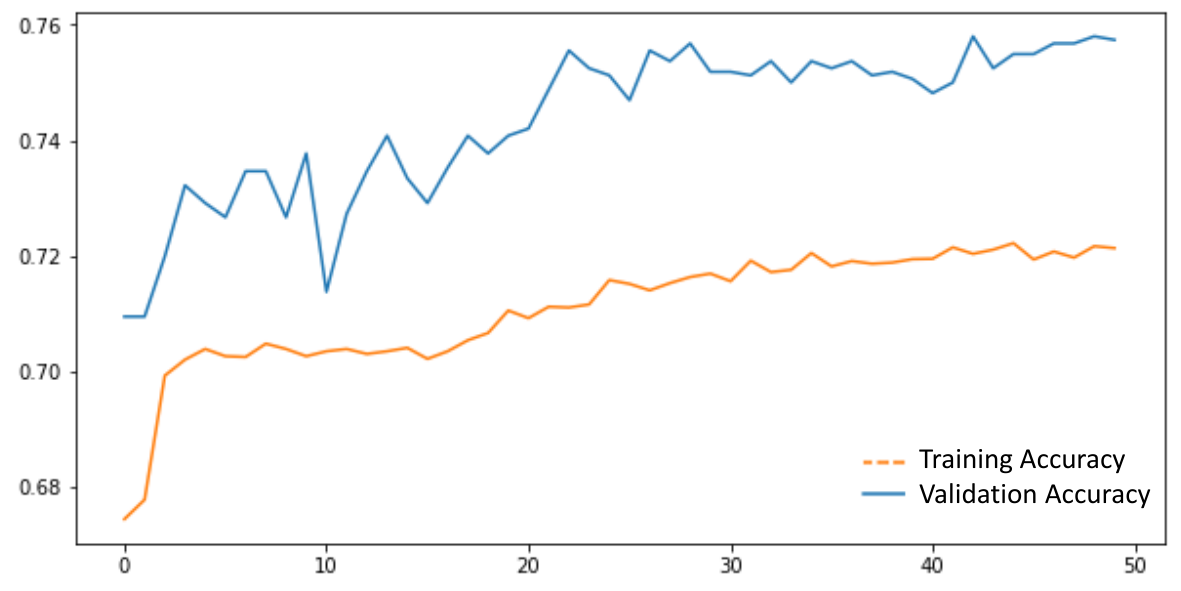}
    \caption{The training and validation accuracy (TPDES state estimation over Time). }
         \label{ACCTRAINING_par}
\end{center}
   
\end{figure}
The training and validation accuracy reach approximately 0.72 and 0.76, respectively. The next step involves testing the FNN. For this purpose, the testing data set is fed to the FNN. For evaluating the FNN's performance, we consider the estimation to be accurate if the true state is assigned the highest probability among all possible states. Consequently, the testing accuracy is almost $73 \%$. Similarly to the previous section, here we compare also with the MBSE approach. In this case we compare the state estimate provided by both approaches at each clock tick and the $MAE$ is computed. The comparison is done over the testing dataset and the resulting $MAE$ is equal to $5.15\%$. Consequently, we can conclude that the FNN developed exhibits good performance. Therefore, let consider a scenario, where the following timed sequence of observations is recorded:
\begin{align*}
    \nu^{t}=(c,1.6)(b,0.4)(c,0.4)(c,1.5)(c,0.9)(b,0.3)(a,0.4)
\end{align*}

Applying our FNN, the results of the state estimation are depicted in Figures \ref{estimate s1 parlly}, \ref{estimate s2 parlly}, \ref{estimate s3 parlly} and \ref{estimate s4 parlly} with blue dots (the results of the state estimation using the MBSE approach are also reported in orange dashed curve).

\begin{figure}[h!]

      \centering
      \includegraphics[scale=0.7]{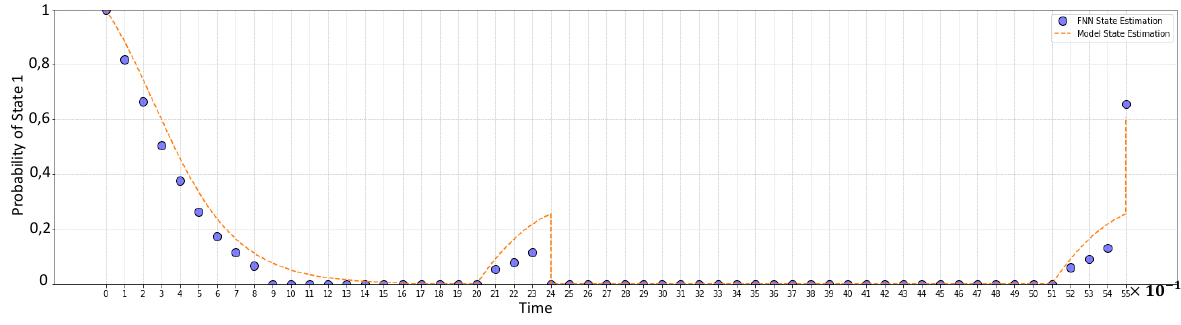}
      \caption{The probability of state $s_1$}
      \label{estimate s1 parlly}
\end{figure}
\begin{figure}[h!]
      \centering
      \includegraphics[scale=0.7]{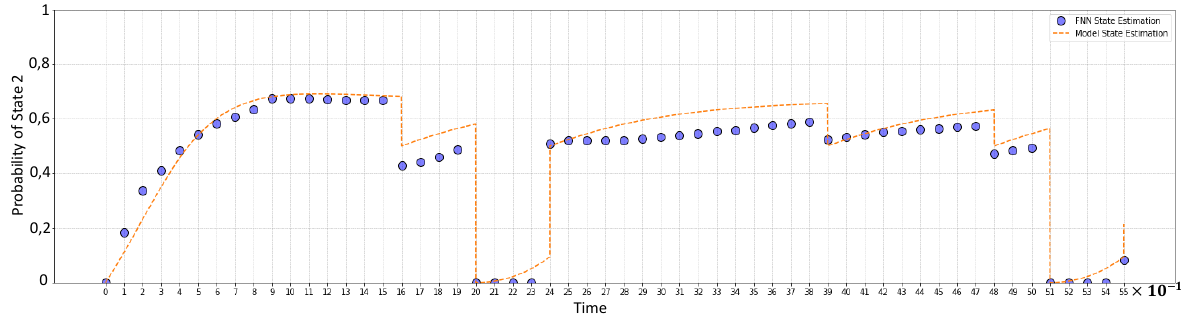}
      \caption{The probability of state $s_2$}
      \label{estimate s2 parlly}
\end{figure}
 \begin{figure}[h!]
      \centering
      \includegraphics[scale=0.7]{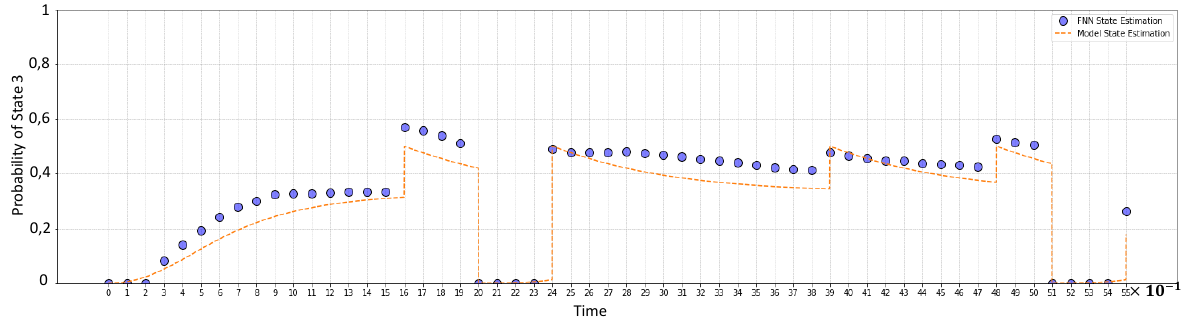}
      \caption{The probability of state $s_3$}
      \label{estimate s3 parlly}
\end{figure}
\begin{figure}[h!]
      \centering
      \includegraphics[scale=0.7]{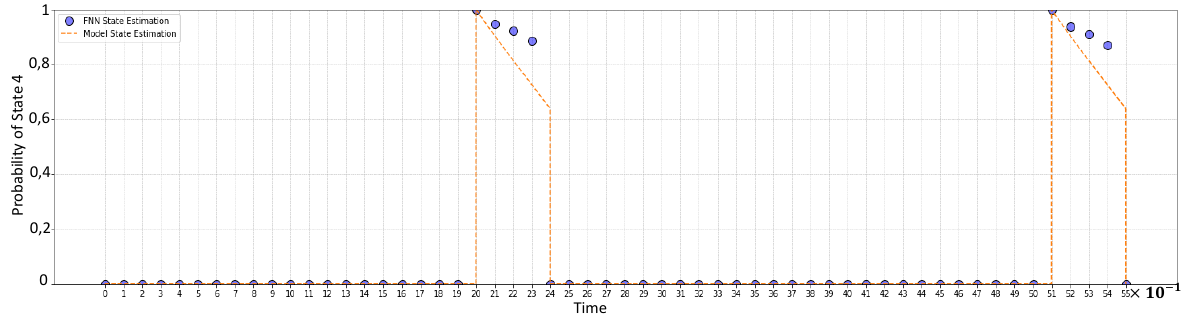}
      \caption{The probability of state $s_4$}
      \label{estimate s4 parlly}
\end{figure}
\end{example}

From these figures, we can remark that the results based on the FNN are very close to those provided by the MBSE highlighting the effectiveness and accuracy of the FNN approach presented in this section.

\section{Conclusions and Perspectives}

In this paper, a comprehensive study on the application of feed-forward neural networks to the state estimation of timed probabilistic discrete event systems is presented. Two main cases are considered: state estimation of TPDES over observations and over time. For each case, the paper outlines the development of a FNN that plays the role of a probabilistic state estimator that benefits from the logical and timed information recorded during the functioning of the system to compute the probability vector of the state of the system. These approaches are particularly significant as they can deal with systems where the occurrence time of events follows different and more complex probability density functions e.g. Weibull or Lognormal distributions. Apart state estimation, these approaches can be also used for various applications such as system's diagnosis, attack detection, online security analysis, opacity, etc. However, they present some limitations, especially, the effectiveness of the FNNs lies on the the data on our disposal, in addition, for complex raw data, a significant effort in data preprocessing may be required, as well as an effective strategy to handle out-of-distribution data should be implemented. Finally, even though the FNN assigns higher probabilities to the states where the system is most likely in, very small, but non-zeros, probabilities are given to other states as well. This issue may lead to difficulties when this estimation is used to decide the system’s correct state. In future works, we aim to study this issue, where some post-processing decision rules will be proposed to refine the state probabilities and to separate good candidates from other ones. 

As perspective of this work, several other directions will be also pursued. Notably, another deep learning tools such as convolutional neural networks and recurrent neural networks will be exploited, in order to find out FNNs that can provide more accurate results. Furthermore, the methods proposed in this paper will be relaxed, so that rather than estimating the state of the system, we will focus on estimating some particular properties of interest that can be represented by sets of states. Finally, practical applications are expected.

\section*{Funding Declaration}

This work has been partially supported by the Region Normandie, France,
Le Havre Seine Metropole (LHSM) RIN ASSAILLANT Project and ANR-22-CE10-0002.

\section*{Acknowledgment}
This preprint has not undergone any post-submission improvements or corrections. The Version of Record of this article is published in Discrete Event Dynamic Systems: Theory and Applications, and is available online at https://doi.org/10.1007/s10626-025-00414-9
  
\bibliography{sn-bibliography}


\begin{thebibliography}{29}
\ifx \bisbn   \undefined \def \bisbn  #1{ISBN #1}\fi
\ifx \binits  \undefined \def \binits#1{#1}\fi
\ifx \bauthor  \undefined \def \bauthor#1{#1}\fi
\ifx \batitle  \undefined \def \batitle#1{#1}\fi
\ifx \bjtitle  \undefined \def \bjtitle#1{#1}\fi
\ifx \bvolume  \undefined \def \bvolume#1{\textbf{#1}}\fi
\ifx \byear  \undefined \def \byear#1{#1}\fi
\ifx \bissue  \undefined \def \bissue#1{#1}\fi
\ifx \bfpage  \undefined \def \bfpage#1{#1}\fi
\ifx \blpage  \undefined \def \blpage #1{#1}\fi
\ifx \burl  \undefined \def \burl#1{\textsf{#1}}\fi
\ifx \doiurl  \undefined \def \doiurl#1{\url{https://doi.org/#1}}\fi
\ifx \betal  \undefined \def \betal{\textit{et al.}}\fi
\ifx \binstitute  \undefined \def \binstitute#1{#1}\fi
\ifx \binstitutionaled  \undefined \def \binstitutionaled#1{#1}\fi
\ifx \bctitle  \undefined \def \bctitle#1{#1}\fi
\ifx \beditor  \undefined \def \beditor#1{#1}\fi
\ifx \bpublisher  \undefined \def \bpublisher#1{#1}\fi
\ifx \bbtitle  \undefined \def \bbtitle#1{#1}\fi
\ifx \bedition  \undefined \def \bedition#1{#1}\fi
\ifx \bseriesno  \undefined \def \bseriesno#1{#1}\fi
\ifx \blocation  \undefined \def \blocation#1{#1}\fi
\ifx \bsertitle  \undefined \def \bsertitle#1{#1}\fi
\ifx \bsnm \undefined \def \bsnm#1{#1}\fi
\ifx \bsuffix \undefined \def \bsuffix#1{#1}\fi
\ifx \bparticle \undefined \def \bparticle#1{#1}\fi
\ifx \barticle \undefined \def \barticle#1{#1}\fi
\bibcommenthead
\ifx \bconfdate \undefined \def \bconfdate #1{#1}\fi
\ifx \botherref \undefined \def \botherref #1{#1}\fi
\ifx \url \undefined \def \url#1{\textsf{#1}}\fi
\ifx \bchapter \undefined \def \bchapter#1{#1}\fi
\ifx \bbook \undefined \def \bbook#1{#1}\fi
\ifx \bcomment \undefined \def \bcomment#1{#1}\fi
\ifx \oauthor \undefined \def \oauthor#1{#1}\fi
\ifx \citeauthoryear \undefined \def \citeauthoryear#1{#1}\fi
\ifx \endbibitem  \undefined \def \endbibitem {}\fi
\ifx \bconflocation  \undefined \def \bconflocation#1{#1}\fi
\ifx \arxivurl  \undefined \def \arxivurl#1{\textsf{#1}}\fi
\csname PreBibitemsHook\endcsname

%%% 1
\bibitem[\protect\citeauthoryear{Shu et~al.}{2008}]{SHU20083054}
\begin{barticle}
\bauthor{\bsnm{Shu}, \binits{S.}},
\bauthor{\bsnm{Lin}, \binits{F.}},
\bauthor{\bsnm{Ying}, \binits{H.}},
\bauthor{\bsnm{Chen}, \binits{X.}}:
\batitle{State estimation and detectability of probabilistic discrete event systems}.
\bjtitle{Automatica}
\bvolume{44}(\bissue{12}),
\bfpage{3054}--\blpage{3060}
(\byear{2008})
\doiurl{10.1016/j.automatica.2008.05.025}
\end{barticle}
\endbibitem

%%% 2
\bibitem[\protect\citeauthoryear{Ramadge}{1986}]{Ram1}
\begin{bchapter}
\bauthor{\bsnm{Ramadge}, \binits{P.J.}}:
\bctitle{Observability of discrete event systems}.
In: \bbtitle{1986 25th IEEE Conference on Decision and Control},
pp. \bfpage{1108}--\blpage{1112}
(\byear{1986}).
\doiurl{10.1109/CDC.1986.267551}
\end{bchapter}
\endbibitem

%%% 3
\bibitem[\protect\citeauthoryear{Caines et~al.}{1988}]{Caines}
\begin{bchapter}
\bauthor{\bsnm{Caines}, \binits{P.E.}},
\bauthor{\bsnm{Greiner}, \binits{R.}},
\bauthor{\bsnm{Wang}, \binits{S.}}:
\bctitle{Dynamical logic observers for finite automata}.
In: \bbtitle{Proceedings of the 27th IEEE Conference on Decision and Control},
pp. \bfpage{226}--\blpage{2331}
(\byear{1988}).
\doiurl{10.1109/CDC.1988.194300}
\end{bchapter}
\endbibitem

%%% 4
\bibitem[\protect\citeauthoryear{Ozveren and Willsky}{1990}]{Ozveren}
\begin{barticle}
\bauthor{\bsnm{Ozveren}, \binits{C.M.}},
\bauthor{\bsnm{Willsky}, \binits{A.S.}}:
\batitle{Observability of discrete event dynamic systems}.
\bjtitle{IEEE Transactions on Automatic Control}
\bvolume{35}(\bissue{7}),
\bfpage{797}--\blpage{806}
(\byear{1990})
\doiurl{10.1109/9.57018}
\end{barticle}
\endbibitem

%%% 5
\bibitem[\protect\citeauthoryear{Gao et~al.}{2020}]{Gao1}
\begin{bchapter}
\bauthor{\bsnm{Gao}, \binits{C.}},
\bauthor{\bsnm{Lefebvre}, \binits{D.}},
\bauthor{\bsnm{Seatzu}, \binits{C.}},
\bauthor{\bsnm{Li}, \binits{Z.}},
\bauthor{\bsnm{Giua}, \binits{A.}}:
\bctitle{A region-based approach for state estimation of timed automata under no event observation}.
In: \bbtitle{2020 25th IEEE International Conference on Emerging Technologies and Factory Automation (ETFA)},
vol. \bseriesno{1},
pp. \bfpage{799}--\blpage{804}
(\byear{2020}).
\doiurl{10.1109/ETFA46521.2020.9211942}
\end{bchapter}
\endbibitem

%%% 6
\bibitem[\protect\citeauthoryear{Li et~al.}{2022}]{Li}
\begin{barticle}
\bauthor{\bsnm{Li}, \binits{J.}},
\bauthor{\bsnm{Lefebvre}, \binits{D.}},
\bauthor{\bsnm{Hadjicostis}, \binits{C.N.}},
\bauthor{\bsnm{Li}, \binits{Z.}}:
\batitle{Observers for a class of timed automata based on elapsed time graphs}.
\bjtitle{IEEE Transactions on Automatic Control}
\bvolume{67}(\bissue{2}),
\bfpage{767}--\blpage{779}
(\byear{2022})
\doiurl{10.1109/TAC.2021.3064542}
\end{barticle}
\endbibitem

%%% 7
\bibitem[\protect\citeauthoryear{Lai et~al.}{2022}]{Lai}
\begin{barticle}
\bauthor{\bsnm{Lai}, \binits{A.}},
\bauthor{\bsnm{Lahaye}, \binits{S.}},
\bauthor{\bsnm{Komenda}, \binits{J.}}:
\batitle{Observer construction for polynomially ambiguous max-plus automata}.
\bjtitle{IEEE Transactions on Automatic Control}
\bvolume{67}(\bissue{3}),
\bfpage{1582}--\blpage{1588}
(\byear{2022})
\doiurl{10.1109/TAC.2021.3069899}
\end{barticle}
\endbibitem

%%% 8
\bibitem[\protect\citeauthoryear{Shu et~al.}{2006}]{Shu1}
\begin{bchapter}
\bauthor{\bsnm{Shu}, \binits{S.}},
\bauthor{\bsnm{Lin}, \binits{F.}},
\bauthor{\bsnm{Ying}, \binits{H.}}:
\bctitle{Detectability of nondeterministic discrete event systems.}
In: \bbtitle{In Proceedings of DCABES},
pp. \bfpage{1040}--\blpage{1044}
(\byear{2006})
\end{bchapter}
\endbibitem

%%% 9
\bibitem[\protect\citeauthoryear{Lefebvre et~al.}{2022a}]{jdeds1}
\begin{botherref}
\oauthor{\bsnm{Lefebvre}, \binits{D.}},
\oauthor{\bsnm{Seatzu}, \binits{C.}},
\oauthor{\bsnm{Hadjicostis}, \binits{C.N.}},
\oauthor{\bsnm{Giua}, \binits{A.}}:
Probabilistic state estimation for labeled continuous time markov models with applications to attack detection
\textbf{32},
65--88
(2022)
\doiurl{10.1007/s10626-021-00348-y}
\end{botherref}
\endbibitem

%%% 10
\bibitem[\protect\citeauthoryear{Lefebvre et~al.}{2022b}]{jdeds2}
\begin{botherref}
\oauthor{\bsnm{Lefebvre}, \binits{D.}},
\oauthor{\bsnm{Seatzu}, \binits{C.}},
\oauthor{\bsnm{Hadjicostis}, \binits{C.N.}},
\oauthor{\bsnm{Giua}, \binits{A.}}:
Correction to: Probabilistic state estimation for labeled continuous time markov models with applications to attack detection
\textbf{32},
539--544
(2022)
\doiurl{10.1007/s10626-022-00364-6}
\end{botherref}
\endbibitem

%%% 11
\bibitem[\protect\citeauthoryear{Lefebvre et~al.}{2023}]{dim3}
\begin{bchapter}
\bauthor{\bsnm{Lefebvre}, \binits{D.}},
\bauthor{\bsnm{Seatzu}, \binits{C.}},
\bauthor{\bsnm{Hadjicostis}, \binits{C.N.}},
\bauthor{\bsnm{Giua}, \binits{A.}}:
\bctitle{Logical and probabilistic aspects of state estimation for markovian systems}.
In: \bbtitle{2023 62nd IEEE Conference on Decision and Control (CDC)},
pp. \bfpage{6929}--\blpage{6935}
(\byear{2023}).
\doiurl{10.1109/CDC49753.2023.10383800}
\end{bchapter}
\endbibitem

%%% 12
\bibitem[\protect\citeauthoryear{Giua and Seatzu}{2002}]{PNSE1}
\begin{barticle}
\bauthor{\bsnm{Giua}, \binits{A.}},
\bauthor{\bsnm{Seatzu}, \binits{C.}}:
\batitle{Observability of place/transition nets}.
\bjtitle{IEEE Transactions on Automatic Control}
\bvolume{47}(\bissue{9}),
\bfpage{1424}--\blpage{1437}
(\byear{2002})
\doiurl{10.1109/TAC.2002.802769}
\end{barticle}
\endbibitem

%%% 13
\bibitem[\protect\citeauthoryear{Giua et~al.}{2007}]{PNSE2}
\begin{barticle}
\bauthor{\bsnm{Giua}, \binits{A.}},
\bauthor{\bsnm{Seatzu}, \binits{C.}},
\bauthor{\bsnm{Corona}, \binits{D.}}:
\batitle{Marking estimation of petri nets with silent transitions}.
\bjtitle{IEEE Transactions on Automatic Control}
\bvolume{52}(\bissue{9}),
\bfpage{1695}--\blpage{1699}
(\byear{2007})
\doiurl{10.1109/TAC.2007.904281}
\end{barticle}
\endbibitem

%%% 14
\bibitem[\protect\citeauthoryear{Wang et~al.}{2011}]{PNSE4}
\begin{barticle}
\bauthor{\bsnm{Wang}, \binits{X.}},
\bauthor{\bsnm{Mahulea}, \binits{C.}},
\bauthor{\bsnm{Júlvez}, \binits{J.}},
\bauthor{\bsnm{Silva}, \binits{M.}}:
\batitle{On state estimation of timed choice-free petri nets}.
\bjtitle{IFAC Proceedings Volumes}
\bvolume{44}(\bissue{1}),
\bfpage{8687}--\blpage{8692}
(\byear{2011})
\doiurl{10.3182/20110828-6-IT-1002.01523} .
\bcomment{18th IFAC World Congress}
\end{barticle}
\endbibitem

%%% 15
\bibitem[\protect\citeauthoryear{Bonhomme}{2015}]{PNSE3}
\begin{barticle}
\bauthor{\bsnm{Bonhomme}, \binits{P.}}:
\batitle{Marking estimation of p-time petri nets with unobservable transitions}.
\bjtitle{IEEE Transactions on Systems, Man, and Cybernetics: Systems}
\bvolume{45}(\bissue{3}),
\bfpage{508}--\blpage{518}
(\byear{2015})
\doiurl{10.1109/TSMC.2014.2353575}
\end{barticle}
\endbibitem

%%% 16
\bibitem[\protect\citeauthoryear{Estrada-Vargas et~al.}{2010}]{DESComp}
\begin{barticle}
\bauthor{\bsnm{Estrada-Vargas}, \binits{A.P.}},
\bauthor{\bsnm{López-Mellado}, \binits{E.}},
\bauthor{\bsnm{Lesage}, \binits{J.-J.}}:
\batitle{A comparative analysis of recent identification approaches for discrete-event systems}.
\bjtitle{Mathematical Problems in Engineering}
\bvolume{2010}(\bissue{1}),
\bfpage{453254}
(\byear{2010})
\doiurl{10.1155/2010/453254}
{\href{https://arxiv.org/abs/https://onlinelibrary.wiley.com/doi/pdf/10.1155/2010/453254}{{https://onlinelibrary.wiley.com/doi/pdf/10.1155/2010/453254}}}
\end{barticle}
\endbibitem

%%% 17
\bibitem[\protect\citeauthoryear{van~der Aalst}{2012}]{processmining}
\begin{botherref}
\oauthor{\bsnm{Aalst}, \binits{W.}}:
Process mining: Overview and opportunities.
ACM Trans. Manage. Inf. Syst.
\textbf{3}(2)
(2012)
\doiurl{10.1145/2229156.2229157}
\end{botherref}
\endbibitem

%%% 18
\bibitem[\protect\citeauthoryear{Saddem and Baptiste}{2022}]{SADDEM}
\begin{barticle}
\bauthor{\bsnm{Saddem}, \binits{R.}},
\bauthor{\bsnm{Baptiste}, \binits{D.}}:
\batitle{Machine learning-based approach for online fault diagnosis of discrete event system}.
\bjtitle{IFAC-PapersOnLine}
\bvolume{55}(\bissue{28}),
\bfpage{337}--\blpage{343}
(\byear{2022})
\doiurl{10.1016/j.ifacol.2022.10.363} .
\bcomment{16th IFAC Workshop on Discrete Event Systems WODES 2022}
\end{barticle}
\endbibitem

%%% 19
\bibitem[\protect\citeauthoryear{Luo et~al.}{2024}]{Reinforcement1}
\begin{barticle}
\bauthor{\bsnm{Luo}, \binits{J.}},
\bauthor{\bsnm{Yi}, \binits{S.}},
\bauthor{\bsnm{Lin}, \binits{Z.}},
\bauthor{\bsnm{Zhang}, \binits{H.}},
\bauthor{\bsnm{Zhou}, \binits{J.}}:
\batitle{Petri-net-based deep reinforcement learning for real-time scheduling of automated manufacturing systems}.
\bjtitle{Journal of Manufacturing Systems}
\bvolume{74},
\bfpage{995}--\blpage{1008}
(\byear{2024})
\doiurl{10.1016/j.jmsy.2024.05.006}
\end{barticle}
\endbibitem

%%% 20
\bibitem[\protect\citeauthoryear{Hu et~al.}{2020}]{Reinforcement2}
\begin{barticle}
\bauthor{\bsnm{Hu}, \binits{L.}},
\bauthor{\bsnm{Liu}, \binits{Z.}},
\bauthor{\bsnm{Hu}, \binits{W.}},
\bauthor{\bsnm{Wang}, \binits{Y.}},
\bauthor{\bsnm{Tan}, \binits{J.}},
\bauthor{\bsnm{Wu}, \binits{F.}}:
\batitle{Petri-net-based dynamic scheduling of flexible manufacturing system via deep reinforcement learning with graph convolutional network}.
\bjtitle{Journal of Manufacturing Systems}
\bvolume{55},
\bfpage{1}--\blpage{14}
(\byear{2020})
\doiurl{10.1016/j.jmsy.2020.02.004}
\end{barticle}
\endbibitem

%%% 21
\bibitem[\protect\citeauthoryear{Chollet}{2021}]{ML1}
\begin{bbook}
\bauthor{\bsnm{Chollet}, \binits{F.}}:
\bbtitle{Deep Learning with Python. Simon and Schuster},
(\byear{2021})
\end{bbook}
\endbibitem

%%% 22
\bibitem[\protect\citeauthoryear{Géron}{2019}]{ML2}
\begin{bbook}
\bauthor{\bsnm{Géron}, \binits{A.}}:
\bbtitle{Hands-on Machine Learning with Scikit-Learn, Keras \& TensorFlow},
(\byear{2019})
\end{bbook}
\endbibitem

%%% 23
\bibitem[\protect\citeauthoryear{Qi et~al.}{2024a}]{Reachability}
\begin{barticle}
\bauthor{\bsnm{Qi}, \binits{H.}},
\bauthor{\bsnm{Guang}, \binits{M.}},
\bauthor{\bsnm{Wang}, \binits{J.}},
\bauthor{\bsnm{Yan}, \binits{C.}},
\bauthor{\bsnm{Jiang}, \binits{C.}}:
\batitle{Probabilistic reachability prediction of unbounded petri nets: A machine learning method}.
\bjtitle{IEEE Transactions on Automation Science and Engineering}
\bvolume{21}(\bissue{3}),
\bfpage{3012}--\blpage{3024}
(\byear{2024})
\doiurl{10.1109/TASE.2023.3272983}
\end{barticle}
\endbibitem

%%% 24
\bibitem[\protect\citeauthoryear{Qi et~al.}{2024b}]{liveness}
\begin{barticle}
\bauthor{\bsnm{Qi}, \binits{H.}},
\bauthor{\bsnm{Wang}, \binits{J.}},
\bauthor{\bsnm{Yan}, \binits{C.}},
\bauthor{\bsnm{Jiang}, \binits{C.}}:
\batitle{The probabilistic liveness decision method of unbounded petri nets based on machine learning}.
\bjtitle{IEEE Transactions on Systems, Man, and Cybernetics: Systems}
\bvolume{54}(\bissue{2}),
\bfpage{1070}--\blpage{1081}
(\byear{2024})
\doiurl{10.1109/TSMC.2023.3323342}
\end{barticle}
\endbibitem

%%% 25
\bibitem[\protect\citeauthoryear{Cassandras and Lafortune}{2007}]{introdes}
\begin{bbook}
\bauthor{\bsnm{Cassandras}, \binits{C.G.}},
\bauthor{\bsnm{Lafortune}, \binits{S.}}:
\bbtitle{Introduction to Discrete Event Systems},
(\byear{2007})
\end{bbook}
\endbibitem

%%% 26
\bibitem[\protect\citeauthoryear{Lefebvre and Hadjicostis}{2022}]{PN33}
\begin{barticle}
\bauthor{\bsnm{Lefebvre}, \binits{D.}},
\bauthor{\bsnm{Hadjicostis}, \binits{C.N.}}:
\batitle{Diagnosability of fault patterns with labeled stochastic petri nets}.
\bjtitle{Information Sciences}
\bvolume{593},
\bfpage{341}--\blpage{363}
(\byear{2022})
\doiurl{10.1016/j.ins.2022.01.061}
\end{barticle}
\endbibitem

%%% 27
\bibitem[\protect\citeauthoryear{Farquha and Gal}{2022}]{OOD}
\begin{bchapter}
\bauthor{\bsnm{Farquha}, \binits{S.}},
\bauthor{\bsnm{Gal}, \binits{Y.}}:
\bctitle{What ‘out-of-distribution’ is and is not}.
(\byear{2022})
\end{bchapter}
\endbibitem

%%% 28
\bibitem[\protect\citeauthoryear{AnalytixLabs}{2024}]{web1}
\begin{botherref}
\oauthor{\bsnm{AnalytixLabs}}:
Activation Functions In Neural Networks: Its Components, Uses \& Types.
Accessed: 2024-07-15
(2024).
\url{https://medium.com/@byanalytixlabs/activation-functions-in-neural-networks-its-components-uses-types-23cfc9a7a6d7}
\end{botherref}
\endbibitem

%%% 29
\bibitem[\protect\citeauthoryear{Radhakrishnan}{2017}]{web2}
\begin{botherref}
\oauthor{\bsnm{Radhakrishnan}, \binits{P.}}:
What are Hyperparameters ? and How to tune the Hyperparameters in a Deep Neural Network?
Accessed: 2024-06-10
(2017).
\url{https://shorturl.at/hnOLE}
\end{botherref}
\endbibitem

\end{thebibliography}

\end{document}